\documentclass[%
 reprint,
superscriptaddress,
groupedaddress,
preprintnumbers,
nobibnotes,
 amsmath,amssymb,
prb,
citeautoscript,
]{revtex4-1}

\newcommand{\sgn}{\mathop{\mathrm{sgn}}}
\newcommand{\ml}{\mathcal{L}}
\newcommand{\mr}{\mathcal{R}}
\newcommand{\mc}{\mathcal}
\newcommand{\tdt}{\tilde{t}}

\usepackage{graphicx}% Include figure files
\usepackage{dcolumn}% Align table columns on decimal point
\usepackage{bm}% bold math
\usepackage[mathlines]{lineno}% Enable numbering of text and display math
\usepackage[rightcaption]{sidecap}

\newcommand{\eps}{\varepsilon}
\newcommand{\vp}{\varphi}

\begin{document}

%\preprint{APS/123-QED}

\title{Coherence recovery mechanisms in quantum Hall edge states.}% Force line breaks with \\

\author{Anna S. Goremykina, Eugene V. Sukhorukov }
\affiliation{%
D\'{e}partement de Physique Th\'{e}orique, Universit\'{e} de Gen\`{e}ve, CH-1211 Gen\`{e}ve 4, Switzerland
}%

\date{\today}
\begin{abstract}
This work is motivated by the puzzling results of the recent experiment [S. Tewari et al., Phys. Rev. B 93, 035420 (2016)], where a robust coherence recovery starting from a certain energy was detected for an electron injected into the quantum Hall edge at the filling factor 2. After passing through a quantum dot the electron then tunnels into the edge with a subsequent propagation towards a symmetric Mach-Zender interferometer, after which the visibility of Aharonov-Bohm (AB) oscillations is measured. According to conventional understanding, its decay  with the increasing energy of the injected electron was expected, which was confirmed theoretically in the bosonization framework. Here we analyze why such a model fails to account for the coherence recovery and demonstrate that the reason is essentially the destructive interference of the two quasiparticles (charge and neutral modes) forming at the edge out of the incoming electron wave packet. This statement is moreover robust with respect to the strength of the Coulomb interaction. We firstly exploit the idea of introducing an imbalance between the quasiparticles, by creating different conditions of propagation for them. It can be done by taking into account either dispersion or dissipation, which indeed results in the partial coherence recovery. The idea of imbalance can also be realized by applying a periodic potential to the arms of interferometer. We discuss such an experiment, which might also shed light on the internal coherence of the two edge excitations. Another scenario relies on the lowering of the energy density of the electron wave packet by the time it arrives at the interferometer in presence of dissipation or dispersion. This energy density is defined by a parameter completely independent of the injected energy, which naturally explains the emergence of a threshold energy in the experiment. 
\end{abstract}
\pacs{Valid PACS appear here}% PACS, the Physics and Astronomy
\maketitle

\section{Introduction}\label{introduction}

The edge excitations of the integer quantum Hall (QH) regime became the basis for the new field of electronic optics in two-dimensional electronic gases, due to their ballistic, one-dimensional and chiral behaviour\cite{review_optics}. At the same time a question about the nature of decoherence  in the QH edge states, important to further quantum information applications, does not yet have a satisfactory answer. Quite a number of the experiments\cite{coh_roulleau, litvin_decoh, tuning_decoh, main_roche, dephase_exp,  dephase_shot_noise, marg_decoh} based on\cite{MZI} the electronic Mach-Zender interferometer (MZI)  tried to shed light on the effects of dephasing and interaction taking place in such systems. Initially, it was demonstrated\cite{MZI} that the coherence of the incoming electron current becomes greatly suppressed with increasing the energy and temperature at the filling factor $\nu=1$. However, even a more striking behavior was revealed in the case of $\nu=2$, where the lobe-structure of the visibility as a function of bias between the arms of the interferometer has been reported \cite{dephase_exp, dephase_shot_noise, lobes_roulleau, litvin_decoh}. Such an effect was attributed \cite{dephase_int} to a strong interaction between the channels, leading to a separation of the spectrum of the edge excitations into the fast charge and slow neutral modes. 

\begin{figure}[h]
\includegraphics[scale=0.4]{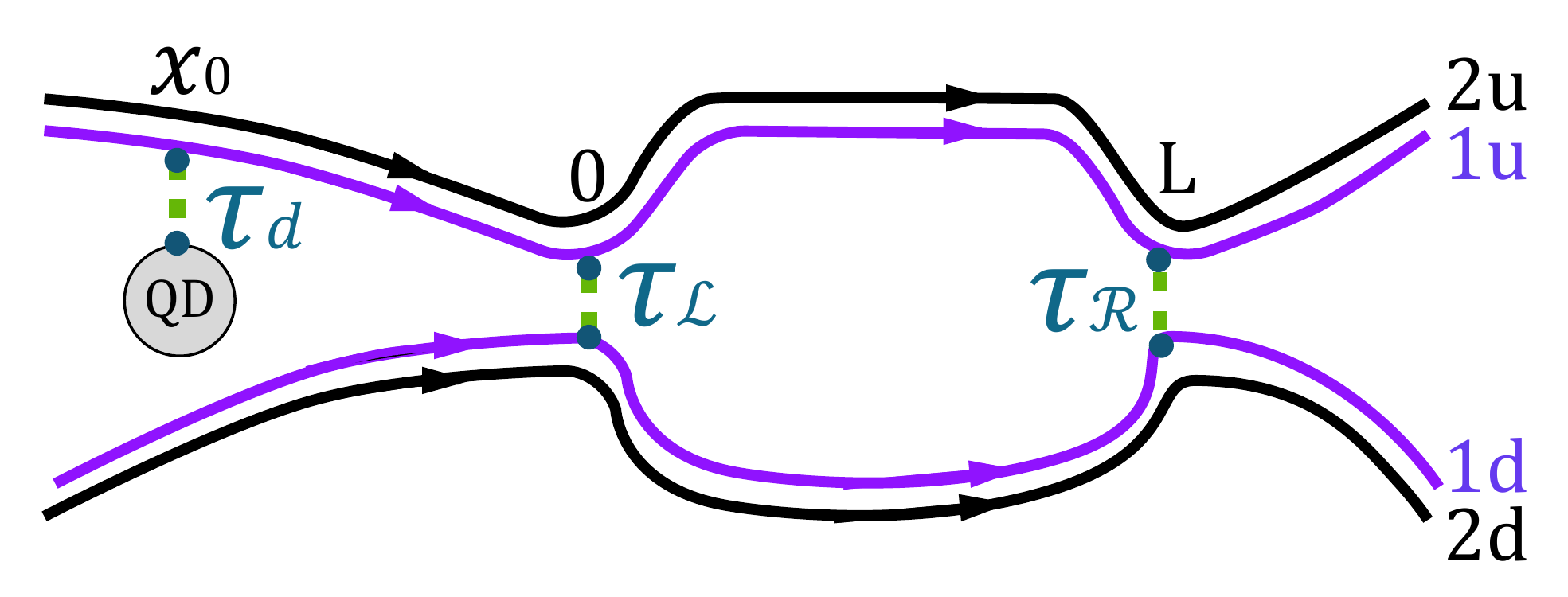} 
\caption{Sketch of the MZ interferometer, comprising the two up and down edge channels of the QH edge system at $\nu=2$. The channels are coupled by the tunnel junctions characterized by the amplitudes $\mc{\tau}_{\mc{L}}$ and $\mc{\tau}_{\mc{R}}$. The electron with the energy $\eps_0$ tunnels to the $1u$ channel from the QD with the amplitude $\mc{\tau}_d$. }
\label{fig:system}
\end{figure}

A later experiment\cite{main_roche} made it possible to concentrate on studying the decoherence of a single electron injected into the edge state. The scheme of the experiment is provided in the Fig.\ \ref{fig:system}. In this set-up, created in the system of QH edge states at the filling factor $\nu=2$, a single-electron wave packet (WP) is injected into one of the channels with the energy defined by the level of the QD, working as an energy filter for an electron passing through. After covering a distance $|x_0|$ of $2.7\mu m$ it eventually arrives at the MZI with the length of $7.2\mu m$ for the both arms. The quantum interference is then analyzed by measuring the oscillations of the current $I$ on the way out of the MZI and plotting subsequently the visibility $V = \frac{I_{max}-I_{min}}{I_{max}+I_{min}}$ as a function of the injected energy. Strikingly, it does not vanish with the energy unlike in the cases mentioned above. Instead, after a short decay, it flattens for the energies larger than $\sim20\mu eV$, with a significant coherence restoration of around $42\%$. In their paper experimentalists argued that such a behavior can be explained by a partial relaxation of the electron WP on its way to the interferometer. Thus, its energy is lowered to the extent when one of the main mechanisms of the decoherence in such a set-up, the inelastic scattering inside the interferometer, would be weak.

An attempt to explain such an outcome was made in [\onlinecite{expln_other}] by taking into account the disorder present in the edge. However, interaction is considered perturbatively in that work and each of the edge states is studied as a Fermi liquid. Another paper [\onlinecite{artur}] adopted an approach from [\onlinecite{dephase_int}], which accounts for the strong Coulomb interaction between the edge channels via the bosonization approach in order to see if such a minimalistic model captures the effect. 
Alas, the visibility was predicted to decay in a power-law manner with increasing the energy of the electron. 

In the present paper -- a logical continuation of the latter work -- we are analyzing what prevented the existence of the visibility plateau in the previous model and describe several possible scenarios of the coherence recovery. Namely, we demonstrate within the initial model that it is the destructive interference of the two quasiparticles originating from the strong interaction between the edge channels that leads to a complete decoherence. Therefore, it is enough to introduce  different conditions guiding their dynamics to prevent such an exact cancellation. Such conditions can already be applied in case of asymmetric interferometer. However, as we point out in the Appendix \ref{app_asymm}, a possible slight asymmetry cannot account for such a significant coherence restoration as witnessed in the experiment. Thus, we move on to more plausible scenarios. A natural assumption would be to include either dissipation or dispersion present in either of the charge or neutral modes. Both of the effects were found to be present\cite{dissip_neutr} at least for the neutral mode, whereas the charge mode has not been studied in the experiment due to its larger excitation energy. At this point it might already seem intuitive that taking either of the effects into account for one of the modes results in the coherence recovery, which is indeed the case as we will demonstrate. On the other hand, there is much more to the physics, when modification of the dynamics is applied to the both modes.

We start our analysis by taking dissipation into account via an additional imaginary term in the velocity of the neutral or the charge modes. In fact, the experiment [\onlinecite{dissip_neutr}] demonstrated more of a quadratic law behavior of the imaginary part of the dispersion relation, which we replace, however, by a linear one for simplicity. We then study the energy distribution function dependence from the injected energy $\eps_0$ and demonstrate the emergence of the energy cut-off $\eps_0\ll 1/\gamma$. Next, we arrive at the following important conclusions:
\begin{itemize}
\item Characteristic time $\gamma$ (i.e. the ``strength'' of dissipation as it is explained below) is directly proportional to $x_0$. Due to the relative analytical simplicity, we concentrate on the case of $x_0\gg L$, which corresponds literally to neglecting the dissipation inside the interferometer. We justify this approach by studying  the limiting situation of $x_0=0$, while including the dissipation inside the MZI in the Appendix \ref{appendix_x0}. Ultimately, we recover the same visibility decay, thus concluding that it is the relaxation of the WP before the interferometer which is more relevant. Hence, the MZI should be considered as a probe of the WP dynamics.
\item Unlike the initial model where the decoherence strength is governed by the parameter $\eps_0\eta$, i.e. the one defining the possible phase space of the scattering states inside the interferometer, it is $\eta/\gamma$ that replaces it in the dissipation case. The latter is derived from the energy loss of the WP and its proportionality to $1/\gamma$ by the time the WP reaches the interferometer.   Throughout the paper we denote $\eta=L/v-L/u$, where $u,v$ are the velocities of the charge and neutral modes respectively and $L$ is a length of the arm of the interferometer.
\item  It follows from the previous point that visibility decays for the energies $\eps_0\gamma\ll 1$ and saturates at the plateau for $\eps_0\gamma\gg 1$. The latter happens due to the inefficiency of the non-elastic scattering inside the interferometer. Particularly, in the case of a strong dissipation $\eta/\gamma\ll 1$ present in both modes, a complete coherence recovery is discovered! For a strong dissipation present in one mode we find a restoration of around $59\%$, while for a small dissipation the restoration appears to be of the order of $\eta/\gamma$.
\item The condition $\gamma\gg\eta$, describing the strong dissipation can be interpreted as a comparison of the effective time broadening of the WP with $\eta$, which explains why non-elastic scattering becomes weak. This argument has direct connections to the dispersion case. Although the energy is conserved, the WP becomes naturally broadened in the space lowering down its energy density. Therefore, we expect the similar physics to take place and calculate only the case of the dispersion (quadratic) present in the neutral mode. Curiously, we arrive at the same result as for the dissipation. 
\item Even though we are not in a position to make precise quantitative predictions we can still estimate the threshold energy $\eps_{thr}$, for instance, in the case of dissipation. Judging roughly from the findings of [\onlinecite{dissip_neutr}], it is of order $\eps_{thr} \sim 40\mu eV$, which is in a good agreement with the experiment. These estimations can be found in Sec.\ \ref{sec_energy_distr_dissip}.
\end{itemize}

What is also intriguing is that despite a total dephasing of the WP in the linear dispersion case, each of the quasiparticles does not decohere on its own. We bring attention to the fact that this statement is robust with respect to the strength of interaction, i.e. supposing the charges of the quasiparticles to differ from $1/2$, which is discussed in the Appendix \ref{appendix_fract}. In order to measure this elusive coherence we propose a following experiment with a remark that we are going to ignore the effects of either dispersion or dissipation in its theoretical description. It is done to single out the bare contribution of a quasiparticle, which nevertheless does not undermine the goal of the experiment itself. Briefly, in addition to the DC current going through a QD one should apply a periodic bias of a small amplitude  between the arms of the MZI, synchronized with the tunneling events. Next, one can separate the two contributions after the MZI and find the visibility of the DC current. What should be found is that it also saturates with the energy, oscillating at the same time, which is exactly the result of the modified time dynamics for one of the modes. Even if the experiment will also show the existence of an additional contribution from the dissipation or dispersion it would be instructive to compare them with such an oscillating part to check our theory.

Finally, we briefly outline the structure of the paper. In Sec. \ref{model} and \ref{transport} we present a formalism to describe a system introduced above as well a transport in it. Then, in Sec. \ref{visib_linear}, we bring up the main result of the paper \cite{artur} for a linear plasmon spectrum and formally analyze the decay of the visibility by studying the large injected energy asymptotics of the interference current. It is next followed by Sec. \ref{sec_corr_func}, where the general approach to finding the correlation functions is discussed. Particularly, we describe in detail, how they are modified by introducing the dissipation via the Fluctuation Dissipation Theorem (FDT). The details of the derivation are provided in the Appendix \ref{appendix_dissip_corr}. Then we proceed to finding the energy distribution function of the WP as well as its energy in Sec. \ref{sec_energy_distr_dissip}. There we provide a clear physical picture of the nature of the coherence recovery and explain how it is essentially connected to the dispersion case. Afterwards, our qualitative predictions are confirmed  by the calculations for the visibility in the presence of either dissipation (Sec.\ \ref{sec_dissip}) or dispersion (Sec.\ \ref{sec_disp}). To wrap up the paper the proposal of an experiment allowing for a periodic coherence recovery is discussed in Sec.\ \ref{sec_period}. Finally, Sec.\ \ref{sec_concl} is devoted to the concluding remarks.

\section{Model and initial state}\label{model}

The basic characteristic of interest is the evolution of the initial state which is a charged QD with a corresponding energy $\eps_0$. We will base its study upon the formalism developed in the previous work [\onlinecite{artur}], where the visibility is studied in the linear spectrum case. We will briefly outline the main steps of that approach and introduce necessary changes in the subsequent Sections \ref{sec_dissip} and \ref{sec_disp}. Note that we work in the units, where $\hbar = c = e = 1$.

The general Hamiltonian describing our system (see Fig.\ \ref{fig:system}) comprising a QD and a MZI has the form
\begin{align}\label{model_hamiltonian}
\mathcal{H} = \mathcal{H}_0 + \mathcal{H}_d + \mathcal{H}_t + \mathcal{H}_{t,d} .
\end{align}
Here $\mathcal{H}_0$ describes the Hamiltonian for the QH edge states for the filling factor $\nu = 2$, $\mathcal{H}_d$ corresponds to the QD's Hamiltonian, while the two left terms take into account the tunneling between the QD and the edge and the tunneling between the ``up'' and ``down'' arms of the interferometer. A standard approach would be to write down such a Hamiltonian in the bosonized form expressing the electron operator $\psi_{nm}(x)$ in terms of a collective bosonic mode $\phi_{nm}(x)$:
\begin{equation}
\psi_{nm}(x) =\frac{1}{\sqrt{2\pi a}} e^{i\phi_{nm}(x)}, \quad  n=1,2; \quad m=u,d,
\end{equation}
where $n$ denotes a particular channel, while $m$ corresponds to the ``up'' or ``down'' part of the interferometer. Also an ultraviolet cut-off $a$ was introduced, whose role is to keep the correct electron anti-commutation relation. 

The fields $\phi_{nm}(x)$ are connected to charge densities $\rho_{nm}(x) = \frac{1}{2\pi}\partial_x\phi_{nm}(x)$ and satisfy the commutation relations $[\phi_{nm}(x), \phi_{n'm'}(y)] = i\pi\delta_{n,n'}\delta_{m,m'}\sgn(x-y)$.
In terms of the new bosonic operators the Hamiltonian $\mathcal{H}_0$ reads:
\begin{equation}
\mathcal{H}_0 = \frac{1}{8\pi^2} \sum_{nn',m} \int dx V_{nn'}(x,y) \partial_x \phi_{nm}(x) \partial_x \phi_{n'm}(x),
\end{equation}
where the matrix $V_{nn'}$  is presented in the form:
\begin{align}
V = \begin{pmatrix}
2\pi v_0 + U & U \\ 
U & 2\pi v_0+ U
\end{pmatrix}.
\end{align}
The term $U > 0$ describes a strong screened Coulomb interaction, while $v_0$ is a Fermi velocity of the edge plasmon in the absence of interaction. Thus, the Hamiltonian $\mathcal{H}_0$ can be easily diagonalized applying a unitary transformation:
\begin{align}\label{model_unit_tr}
\phi_{1m} = \frac{1}{\sqrt{2}}\left(\vp_{1m}+\vp_{2m}\right), \quad \phi_{2m} = \frac{1}{\sqrt{2}}\left(\vp_{1m}-\vp_{2m}\right).
\end{align}
The new fields $\vp_{1m}$ and $\vp_{2m}$ are called respectively charge and neutral modes according to their ``physical sense''. In their basis $\mathcal{H}_0$ acquires a form:
\begin{equation}\label{model_diag_ham}
\mathcal{H}_0 = \frac{1}{4\pi} \int dx \sum_{n,m}v_n \lbrace \partial_x\vp_{nm} \rbrace^2,
\end{equation}
with the charge mode velocity $v_1 \equiv u = v_0 + U/\pi$ being much larger than the velocity of the dipole mode $v_2 \equiv v_0$ due to the strong Coulomb interaction $U\gg v_0$. Note that we will mention below and discuss in the Appendix how this condition can be relaxed and outline the consequences.

Moving to the Hamiltonian for the QD, we represent it in terms of the electron annihilation operator $d$, which gives
\begin{equation}
\mathcal{H}_d = \eps_0 d^{\dagger} d,
\end{equation}
where $\eps_0$ is the energy level of the charged QD. Next, the tunneling term $\mathcal{H}_{t,d}$ between the dot and the channel $1u$  in the Hamiltonian is presented as:
\begin{align}
\mathcal{H}_{t,d}=\frac{\tau_d}{\sqrt{2\pi a}} e^{-i \phi_{1u}(x_0)}d +h.c.
\end{align}
Finally, the tunneling Hamiltonian $\mathcal{H}_t$ between the $1d$ and $1u$ channels of the interferometer has the form
\begin{equation}\label{model_mz_tunneling}
\mathcal{H}_t = \sum_{j=\mathcal{L},\mathcal{R}} \frac{\tau_j}{2\pi a} e^{i\phi_{1u}(x_j)}e^{-i\phi_{1d}(x_j)}+ h.c, 
\end{equation}
where $x_{\mathcal{L}}=0$ and $x_{\mathcal{R}}=L$.

Having defined the Hamiltonian, we move to studying the evolution of a single-particle state injected into the QH edge from a QD. To account for tunneling into the edge, described by $\tau_d$, non-perturbatively, the evolved single-particle state can be presented in a form of a wave packet (WP) of a certain width $\Gamma$, whose dynamics is governed by the Hamiltonian 
\begin{equation}\label{ham_tot}
{H}_{tot} = \mc{H}_0 +\mc{H}_d + \mc{H}_{t,d}.
\end{equation} 
Such a packet will then describe a single-particle state at large times with an empty QD. The next idea is to construct an auxiliary initial state for a new system {\it without} a QD, whose evolution will still describe the old system comprising the edge and the QD.  For that, the WP must be evolved back in time with $\mc{H}_0$. In the simplest case of $\nu = 1$ the result\cite{artur} reads
\begin{equation}\label{model_init_state}
\vert\Psi\rangle_{in} = \frac{\vert\tau_d\vert}{v_0}\int_{-\infty}^{x_0}dx e^{i(\eps_0-i \Gamma)(x-x_0)/v_0}e^{i\phi(x)}\vert \Omega\rangle,
\end{equation}
where $\Gamma = \vert \tau_d\vert^2/2v_0$ is the QD level width and $\vert \Omega\rangle$ stands for the ground state of the system. This state, being evolved with the edge Hamiltonian $\mc{H}_0$, coincides at time $t = T \gg \Gamma^{-1}$ with the state injected from a QD and evolved with the total Hamiltonian \eqref{ham_tot}. Thus, it allows reformulating the problem: instead of studying the evolution of an injected electron, one studies the dynamics of an auxiliary initial state of the system without a QD.

But what happens if an electron tunnels into the edge of $\nu=2$ with a non negligible interaction between the channels? We still may use the same initial state with a few modifications. Firstly, an obvious replacement of $\phi(x)$ by $\phi_{1u}(x)$ is in order. Secondly, 
this state should be considered as the one coming in from the free-fermion region $x \leq x_0$ and propagating then into the region $x>x_0$, where there is interaction between the two channels.\footnote{Separation of the space into a ``free'' region and the one with interactions between the edge channels and/or other effects is somewhat artificial and originates from the choice of representing the initial state. Otherwise, we would have to account for higher order tunneling processes into the edge.} Next, one can add to that dissipation or dispersion. Eventually, the only quantities of interest are either the correlators for the fields inside the interferometer or the correlators between the free bosonic field $\phi(x)$, present in the WP \eqref{model_init_state}, and the charge or neutral modes in the region $x>x_0$. The contribution from those two modes can be treated separately, as if there was only a channel with no effective interaction region at $x<x_0$, and a region where an interaction is introduced, which changes the fields velocity and/or where the dissipation or dispersion are ``turned on''. We will return to this matter later on, when we get down to the calculations for different regimes.

\section{Transport through the interferometer}\label{transport}

To  find the visibility we first define currents through the interferometer, measured in the $1d$ channel, in the first order in tunneling amplitudes $\tau_L$ and $\tau_R$ in \eqref{model_mz_tunneling} for the left and right junctions. The general expression for the current has the following structure\cite{artur}
\begin{equation}\label{interf_current_general}
I(t) = \sum_{j,j'=\mathcal{L},\mathcal{R}}\langle I_{jj'}(t)\rangle,
\end{equation}
where each term is of the form
\begin{equation}\label{current_def}
I_{jj'}(t)=-\int_{-\infty}^t dt' I_{jj'}(t,t'),
\end{equation}
such that 
\begin{align}
I_{jj'}(t,t')&=\left[A^{\dagger}_{j'}(t'), A_{j}(t)\right] + \left[A^{\dagger}_{j'}(t), A_{j}(t')\right],\\
A_j(t)&= \frac{\tau_j}{2\pi a} e^{i\phi_{1u}(x_j)}e^{-i\phi_{1d}(x_j)}.
\end{align}
The averaging in \eqref{interf_current_general} is taken over the initial state \eqref{model_init_state} with two remarks. Firstly, the ground state of the system is a product state of the ``up'' and ``down'' subsystems: $\vert \Omega\rangle = \vert \Omega_u\rangle \vert \Omega_d\rangle$, so that each $\vert\Omega_{u,d} \rangle= \vert 0\rangle_1\vert 0\rangle_2$ is the ground state of the two channels formed by the ground states of the charge and neutral modes. The terms $\langle I_{\mathcal{L}\mathcal{L}}\rangle$ and $\langle I_{\mathcal{R}\mathcal{R}}\rangle$ are the direct currents at the left and right tunnel junctions respectively, while the interference currents are represented by $\langle I_{\mathcal{L}\mathcal{R}}\rangle$ and $\langle I_{\mathcal{R}\mathcal{L}}\rangle$. These currents define the corresponding direct and interference charges 
\begin{align}
Q_{dir} &= \int dt\left[\langle I_{\mathcal{L}\mathcal{L}}(t)\rangle + \langle I_{\mathcal{R}\mathcal{R}}(t)\rangle\right], \\
Q_{int} &= 2\int dt \mathrm{Re}\lbrace\langle I_{\mathcal{L}\mathcal{R}}(t)\rangle\rbrace\label{interf_current_charge_int},
\end{align}
which in turns determine the visibility $V$:
\begin{equation}\label{visibilit_def}
V = \frac{Q_{max}-Q_{min}}{Q_{max}+Q_{min}} = \frac{\vert Q_{int}\vert}{Q_{dir}}.
\end{equation}
The charges $Q_{max}$ and $Q_{min}$ in the above expression correspond to the maximum and minimum values of the charge $Q = Q_{dir} + Q_{int}$ transmitted through the interferometer, which oscillates as a function of the AB phase $\vp_{AB}=\mathrm{Arg}(\tau_{\mathcal{L}}\tau^{*}_{\mathcal{R}})$ in $Q_{int}=\vert Q_{int}\vert \cos(\vp_{AB}+\theta)$, where $\theta$ is an additional scattering phase shift. 
Throughout the paper we will be using a convenient notation $\mc{Q}\equiv \int_{-\infty}^{\infty} dt \langle I_{\mathcal{L}\mathcal{R}}(t)\rangle$.
%The direct currents do not depend on dissipation or dispersion, which appear in the plasmon velocity, and, therefore, only contribute to the correlation functions found at the different space coordinates. Hence, the direct charge, defined as an integral over time of the sum of the two direct currents simply reads\cite{artur}
%\begin{equation}\label{interf_current_direct}
%Q_{dir} =  \frac{\vert\tau_{\mathcal{L}}\vert^2+\vert\tau_{\mathcal{R}}\vert^2}{uv}.
%\end{equation}
%The interference charge on the other hand is more challenging to find, as it also contains the key to understanding the nature of the effect that we are studying.
\section{Visibility: case of linear spectrum of plasmons}\label{visib_linear}
Before we move to introducing dissipation or dispersion effects, it is instructive to understand what prevents the coherence recovery of visibility, if none of those phenomenon is taken into account. In what follows we are going to analyze previously obtained result for the visibility, looking at it, however, from a new point of view.

Firstly, direct charge acquires a trivial form:
\begin{align}
Q_{dir} = \frac{\vert\tau_{\ml}\vert^2+\vert\tau_{\mr}\vert^2}{uv}.
\end{align}
We note that the direct charge or current are the local characteristics, that originate from the local correlation functions either at the point $x = 0$ or $x=L$. As a result this quantity does not depend on the distance $x_0$ between the QD and the left tunneling junction. The interference current and hence the associated charge has a trickier structure, which reads
\begin{align}\label{interf_current}
\langle I_{\ml\mr}(t)\rangle =
&-\frac{\Gamma \tau_{\ml}\tau^{*}_{\mr}}{2\pi^2 uv\eta}\iint_{-\infty}^0 d\tau d\tau' \frac{e^{i\eps_0 (\tau-\tau')+\Gamma(\tau+\tau')}}{\tau-\tau'-i\delta}\nonumber\\
&\times\left[F(u,v)-F(v,u)\right],
\end{align}
where 
\begin{align}
F(u,v) = &\frac{\sqrt{-i(\tau'+t+x_0/v-L/v)+\gamma}}{\sqrt{-i(\tau'+t+x_0/v-L/u)+\gamma}}\nonumber\\
\times&\frac{\sqrt{i(\tau+t+x_0/v-L/u)+\gamma}}{\sqrt{i(\tau+t+x_0/v-L/v)+\gamma}}.\label{interf_current_F}
\end{align}
Here $\eta = L/v-L/u$ is a flight time difference between the two modes along the interferometer, mentioned in Sec. \ref{introduction}. In the original paper [\onlinecite{artur}], the expression \eqref{interf_current} was carefully transformed to a simpler form to perform further both analytical and numerical calculations. However, as we are only interested in whether the visibility saturates or decays, we are going to find the current asymptotics for a large initial energy $\eps_0\to\infty$.

In such a limit the biggest contribution to the integral comes from $F(u,v)-F(v,u)$ taken at $\tau=\tau'$. Recalling that in order to obtain the interference charge \eqref{interf_current_charge_int} one needs to integrate the current \eqref{interf_current} over $t$, we shift $t$ by $\tau+ x_0/v-L/v$ and arrive at:
\begin{align}
Q_{int} \propto 
\int dt &\left( \sqrt{\frac{t+i\gamma}{t-i\gamma}}\sqrt{\frac{t+\eta-i\gamma}{t+\eta+i\gamma}}\right.\nonumber\\
&\left.- \sqrt{\frac{t-i\gamma}{t+i\gamma}}\sqrt{\frac{t+\eta+i\gamma}{t+\eta-i\gamma}}\right) = 0,\label{charge_asympt_simple_1}
\end{align}
which is justified by the relation $\sqrt{\frac{t+i\gamma}{t-i\gamma}} = \sgn(t)$ when $\gamma\to 0.$ Therefore, there are two contributions from the fractional edge excitations (quasiparticles) $F(u,v)$ and $F(v,u)$ into the interference current that exactly cancel each other as energy $\eps_0$ is increased, meaning the decay of the visibility. Remarkably, each quasiparticle does not decohere on its own and it is their total contribution to the current that leads to vanishing of the visibility. Interestingly, the effect is robust with respect to the strength of interaction when the fractional charges of the quasiparticles differ from $1/2$, which is discussed in the Appendix \ref{appendix_fract}.
We, thus, conclude that the exact cancellation might be prevented by introducing different propagating conditions for the two modes.
Remaining within the current model, we study the effect of a possible slight asymmetry in the interferometer (Appendix \ref{app_asymm}), which provides such conditions. However, we conclude that within the particular experiment, which is stated to be conducted with a symmetric interferometer, a slight asymmetry contribution can not playing a decisive role in the large coherence recovery effect.

Finally, we note that the case of free fermions for $\nu=1$ can be easily recovered from the above formulas by taking $u=v$, resulting in the following visibility:
\begin{equation}\label{visibilit_free}
V_0= 2\frac{\vert\tau_{\ml}\tau_{\mr}\vert}{\vert\tau_{\ml}\vert^2+\vert\tau_{\mr}\vert^2}.
\end{equation}

\section{Correlation functions}\label{sec_corr_func}

\subsection{General approach}
Expression for the current \eqref{interf_current_general} consists of various electron correlation functions, which can be separated into the ``up'' and ``down'' groups, for the operators in the corresponding arms of the interferometer. As we are supposing the size $L\ll |x_0|$ of the interferometer to be small and hence neglecting effects of dispersion or dissipation inside it, correlators for the ``down'' electron operators are of the following form:
\begin{align}
\langle\Omega_d\vert\Psi_{1d}^{\dagger}(x,t)&\Psi_{1d}^{\dagger}(y,t')\vert\Omega_d\rangle =\nonumber\\
&\frac{1}{2\pi a}\frac{1}{\sqrt{i(y-vt')-i(x-vt)+\delta}}\nonumber\\
&\times\frac{1}{\sqrt{i(y-ut')-i(x-ut)+\delta}},
\end{align}
where $\delta^{-1}$ is the energy ultra-violet cut-off.
Such a result follows directly from the expansion of the bosonic field $\phi(x,t)$ into the eigenstates of a free Hamiltonian \eqref{model_diag_ham} with a particular velocity $v$:
\begin{equation}\label{phi_exp}
\phi(x,t) = \int_{-\infty}^{\infty}\frac{dk}{\sqrt{k}} \lbrace e^{ikx-ikvt}\hat{a}_k +h.c.\rbrace
\end{equation}
and the relation $\langle\hat{a}_k\hat{a}_k^{\dagger}\rangle = \theta(k)$ for the boson creation and annihilation operators at zero temperature.
So averaging ``down'' operators with the initial state \eqref{model_init_state} also becomes trivial. However, the ``up'' group includes correlators between the operators forming the initial wave-packet at $x<x_0$ and the ones in the region $x>x_0$. The way such correlators have been treated\cite{artur} greatly used translational symmetry both in time and space, which is only possible for the case of linear spectrum. Thus, we are going to need a more general approach. One could start with a specific 1D equation of motion for the bosonic field. For instance, the following equation
\begin{equation}
\dot{\phi}(x,t) +\lbrace v_0 + [v-v_0]\theta(x-x_0)\rbrace\partial_x \phi(x,t) = 0
\end{equation}
allows finding the bosonic field with a velocity $v$ in the region $x>x_0$ in terms of the field $\phi(x<x_0,t)$. It then gives all necessary correlations for the linear spectrum case.\footnote{With such an approach one must not forget that the density of states proportional to $1/v_0$ in the free region changes to $1/\sqrt{uv}$ in the interaction region. Thus, to keep electron operator continuous in space one must demand $v_0 = \sqrt{uv}$.} To include the dispersion one needs to slightly modify the above equation and follow the same procedure. On the other hand, such a method fails to describe a dissipation, as it is impossible then to use the expansion over the Hamiltonian eigenstates in the region $x>x_0$. Therefore, we are going to handle it making use of the FDT, to which we devote the discussion below.

\subsection{Correlation functions for dissipation}

The linear response theory provides means of determining the correlator $\langle\phi(y<x_0)\phi(x>x_0,t)\rangle$ directly by connecting it to the response function to perturbation (FDT). In its turn, such a response function is found from the equation of motion for $\phi(x,t)$ with the source defined by the form of the perturbation. Such a method has a great advantage as it does not require any knowledge of the Hamiltonian, but the spectrum. It is therefore a perfect approach to describe a dissipative behavior.  The dissipation itself will be accounted for phenomenologically by adding an imaginary term $iv'$ to the plasmon velocity $v$ in the region $x>x_0$, which is justified by the findings of the recent experiment.\cite{dissip_neutr} For the sake of simplicity we suppose the velocity to be constant along $x$.

To proceed with our scheme, we first choose a convenient form of perturbation: $V(x,t) \equiv \delta(x-y) V(t)$, where $y<x_0$ is a particular coordinate. Then, the corresponding additional term to the Hamiltonian reads  $\mc{H}_1 = -\int_{-\infty}^{\infty}\rho(x,t)V(x,t)= -\rho(y,t)V(t)$, where $\rho(y,t)$ is a charge density. According to the Kubo formula we get then:
\begin{align}
&\langle\delta\rho(x,t)\rangle = i\int_{-\infty}^0 d\tau V(t+\tau)K(y,x,\tau),\\
&K(y,x,\tau) = \langle[\rho(y,\tau),\rho(x,0)]\rangle.
\end{align}
Hence, the response function is of the form
\begin{align}
\mc{G}(y,x,\omega) = \frac{\delta\rho_{\omega}}{\delta V_{\omega}} = i\int_{-\infty}^0dt e^{i\omega t}K(y,x,t).
\end{align}
Next, one can easily demonstrate that 
\begin{align}\label{dissip_fdt}
S_{\rho}(k,\omega) = -2\theta(\omega)\mathrm{Im}\mc{G}(k,-\omega),
\end{align}
where $S_{\rho}(k,\omega) = \int dx dt e^{-ik(x-y)+i\omega t}\langle\rho(y,t),\rho(x,0)\rangle$.

The second step would be to find the response function to the perturbation from the following equation of motion with the source, generated by the form of the spectrum and the dissipation:
\begin{align}
-i\omega \phi(x,\omega) +\lbrace v+iv'\theta(x-x_0)\rbrace\partial_x\phi(x,\omega)&\nonumber \\
= V_{\omega}\delta(x-y)&.\label{dissip_eq}
\end{align}
Its solution in the frequency domain acquires the form:
\begin{align}
&\phi(y,x,x_0,\omega) =\frac{V_{\omega}}{v}e^{-i\frac{\omega y}{v}}\theta(x-y)\times\nonumber\\
&\lbrace\theta(x_0-x)e^{i\frac{\omega x}{v}}
+\theta(x-x_0)e^{i\frac{\omega}{v+iv'}(x-x_0)+i\frac{\omega x_0}{v}}\rbrace.\label{dissip_field_sol_1}
\end{align}
Note, that the form of solution implies that $v'<0$, to ensure convergence in the region $x>x_0$.
Moving then to the $k$ domain for all the space variables and finding the response function (see Appendix \ref{appendix_dissip_corr} for the details of calculation), we arrive at the following result:
\begin{align}
\langle\phi(y)\phi(x,t)&-\phi(y)\phi(x,0)\rangle\nonumber\\
&= \ln\frac{-i(\frac{y-x_0}{v}+\frac{x_0-x}{v-iv'})}{-i(t+\frac{y-x_0}{v}+\frac{x_0-x}{v-iv'})}.\label{dissip_correlator}
\end{align}
To introduce a different velocity $v_2$ in the region $x>x_0$ one can simply replace $v-iv'$ by $v_2-iv'$, which can be reached by adding an additional term into \eqref{dissip_eq}. We avoided that to concentrate on the dissipation effect.

The expression \eqref{dissip_correlator} has a very clear structure and one easily restores the correlation functions for a simple linear spectrum by putting $v'$ to 0. Hence, the only difference between the two cases is the renormalization of the velocity for $x>x_0$: $\tilde{v}^{-1} = \frac{v}{v^2+v'^2}$, but more importantly a {\it finite} shift 
\begin{equation}\label{gamma}
\gamma = \frac{x_0-x}{v^2+v'^2}v' 
\end{equation}
of the logarithm branch point into the complex half-plane, which was infinitesimally small before. We are interested in the correlators like \eqref{dissip_correlator} for $x=0$ or $x=L$. Consequently, $\gamma$ acquires positive values and is proportional to the distance between the QD and the interferometer, which we assume to be the largest distance scale in our problem. One needs to remember that such a correlator goes into the exponent with a prefactor of $1/2$ as $\phi(x)$ is either a neutral or charge field. 

\section{Energy distribution function: threshold emergence}\label{sec_energy_distr_dissip}

In this section we concentrate mostly on the energy distribution function in presence of dissipation. Being relatively simple to account for on the calculations level, it also allows to grasp the physics quickly. Dispersion, on the other hand, is harder to deal with analytically, which nevertheless is not a problem on a qualitative level since the behaviour of the WP in its presence is quite clear: it broadens in space and cools down locally. We, therefore, linger on the case of dissipation and return to its comparison with the effect of dispersion in the end.

Dissipation may be present both in the regions outside and inside the interferometer. First and foremost we would like to answer the question: which contribution is more important? Taking dissipation into account everywhere leads to cumbersome expressions without a clear analytical structure. So let us concentrate on two cases: $x_0 = 0$, so that the tunneling from the dot takes place directly into the interferometer, and $|x_0|\gg L$. 
Therefore, in the first case we only study the effect of dissipation inside the interferometer. We provide a careful discussion of this limit in the Appendix \ref{appendix_x0}, where we demonstrate, that for $\eps_0\to \infty$ the interference current vanishes. On the contrary, as it will be seen from the current Section, $|x_0|\gg L$ case shows that it is the dissipation outside the interferometer that leads to the coherence recovery. Thus, we expect a crossover between the two cases and a partial coherence recovery for finite $x_0$. In this sense interferometer should be considered as a probe of the physics happening with the WP on its way to it.

Before moving to any calculations for the current we start with discussing the energy distribution function of the injected electron in a particular channel $n={1u; 2u}$:
\begin{equation}
f_n (\eps,x) = \int dt \Delta W_n(\eps,t,x),
\end{equation}
described by the Wigner function\cite{dario}
\begin{align}
&W_n (\eps,t,x) = \int dz \frac{e^{-i\eps z}}{2\pi}\langle\Psi^{\dagger}_n(x,t+z/2)\Psi_n(x,t-z/2)\rangle,\\
&\Delta W_n(\eps,t,x) = W_n(\eps,t,x) - W^{FS}_n(\eps,t,x),
\end{align}
where $W^{FS}_n(\eps,t,x)$ is a Fermi Sea contribution. We concentrate on $f_{1u}(\eps,x=0)$, whose exact form is presented below: 
\begin{align}
f_{1u}(\eps,0) &= -\frac{\Gamma}{4\pi^3}\int_{-\infty}^{\infty}dt\int_{-\infty}^{\infty}dz \frac{e^{-i\eps z}}{z-i\delta}\nonumber\\
&\times\iint_{-\infty}^0 d\tau d\tau' \frac{e^{i\eps_0(\tau-\tau')+\Gamma(\tau+\tau')}}{\tau-\tau'-i\delta}\nonumber\\
&\times\{\chi_v(t,z,\tau,\tau')\chi_u(t,z,\tau,\tau') - 1\},\label{energy_distr_1}
\end{align}
where 
\begin{align}
\chi_v(t,z,\tau,\tau') &= \frac{\sqrt{-i(t+z/2+\tau+x_0/v)+\gamma}}{\sqrt{-i(t-z/2+\tau+x_0/v)+\gamma}}\nonumber\\
&\times\frac{\sqrt{i(t-z/2+\tau'+x_0/v)+\gamma}}{\sqrt{i(t+z/2+\tau'+x_0/v)+\gamma}}\label{energy_distr_chi_1}
\end{align}
and $\chi_u$ is obtained from $\chi_v$ by replacing $v$ by $u$ and $\gamma$ by another one if necessary.

Let us suppose that dissipation is present only in the neutral mode, so that $\gamma$ in \eqref{energy_distr_chi_1} is finite and is infinitesimally small in the charge mode, represented by $\chi_u$. To analyze the cumbersome expression for the energy distribution function we consider ``weak'' dissipation in a sense that $\gamma\ll x_0/u-x_0/v$. In such a case $f_{1u}(\eps)$ can be easily treated\footnote{Note that while in the case of a linear spectrum, it was the two QPs, whose destructive interference lead to the complete decoherence, ``turning'' on the dissipation smears their distinctive characteristics, by destructing and cooling them down. Thus,  we were able to separate them in $\chi_v\chi_v-1$ in \eqref{energy_distr_1}only for a ``small'' dissipation.} by a separation of the contributions from the neutral and charge modes $\chi_v\chi_u-1 = \chi_v-1 + (\chi_u-1)$, so that $f_{1u} = f_v+f_u$. The behaviour of $f_u$ is known\cite{artur} and is governed by the law $f_u(\eps\to\infty) \sim 1/\eps$ in the limit $\eps_0\to\infty$. However, the contribution influenced by the dissipation acquires an energy cut-off of the order $1/\gamma$ in the same limit:
\begin{equation}
f_v(\eps) \sim \frac{1}{\eps}e^{-2\eps\gamma},
\end{equation}
which can be verified by an asymptotic integration of
\begin{align}\label{energy_distr_neutral}
 f_v \propto \iint_{-\infty}^{\infty}dtdz \left(\frac{e^{-i\eps z}}{z-i\delta}\sqrt{\frac{t+i\gamma}{t-i\gamma}}\sqrt{\frac{t+z-i\gamma}{t+z+i\gamma}}-1\right)
\end{align}
derived from Eq.\ \eqref{energy_distr_1} for a large initial energy. Alternatively, one can study a more convenient value $\partial f_{v}(\eps)/\partial\eps$, whose exact form can be simply revealed:
\begin{align}
\frac{\partial f_{v}}{\partial \eps} &\sim - \frac{\gamma^2}{\pi^2}\lbrace K_1(\gamma\vert\eps\vert)\mathrm{\sgn}(\eps)+K_0(\gamma\vert\eps\vert)\rbrace^2 \nonumber\\
&\sim -\frac{2\gamma}{\pi\eps}e^{-2\gamma\eps} \quad \text{for}\quad \eps\to\infty\label{energy_distr_result_1},
\end{align}
where $\mc{K}_{0,1}$ is the modified Bessel function of the second kind. The plot of the above asymptotics for $\frac{\partial f_{v}}{\partial \eps}$ is presented in the Fig.\ \ref{fig:dfde} below.
\begin{figure}[h]
\includegraphics[scale=0.4]{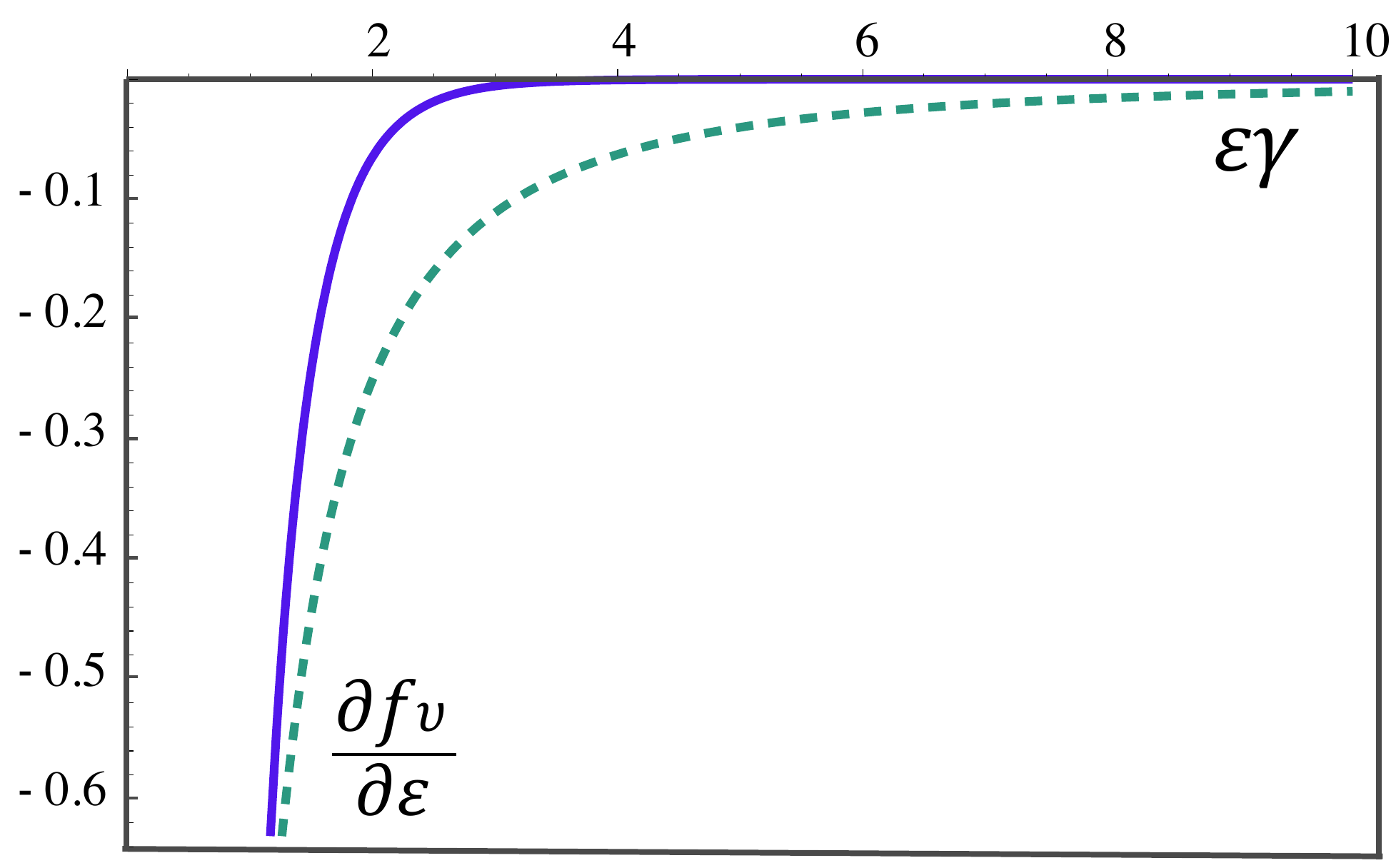} 
\caption{Comparison of derivatives for the energy distribution functions in two cases: $1)$ in presence of dissipation (solid line), according to \eqref{energy_distr_result_1}  and $2)$ the simplest case of linear dispersion\citep{artur} (dashed line), where $\partial f_v/\partial\eps \propto -1/\eps^2$. These plots reflect the existence of the energy cut-off $\sim1/2\gamma$ in the former situation, which is found from the assumption that the initial WP energy is large enough $\eps_0 \gamma\gg 1$. Thus, by the time the WP reaches the interferometer it loses its energy, which is defined by the strength of dissipation $\gamma$.}
\label{fig:dfde}
\end{figure}

Such a behaviour implies that for $\eps_0\gamma\gg 1$ it is the cut-off $\gamma^{-1}$ that defines the energy of the mode by the time it reaches the interferometer. Namely, the latter becomes proportional to $\gamma^{-1}$ as it is shown in Appendix \ref{app_energy_dissip}. On the other hand, for energies below the cut-off $\eps_0 \gamma< 1$, the injected state remembers its initial conditions.
Hence, such a threshold phenomena will be reflected in the visibility behaviour in a form of a robust coherence plateau from a certain point of order $\gamma^{-1}$. 

At this point it is instructive to check if the estimation for the threshold energy $\eps_{thr} \sim \hbar/2\gamma$ gives a reasonable number. For that we use the definition \eqref{gamma} of $\gamma$, with $v'$ denoting the imaginary contribution to the velocity. We then roughly estimate the velocities $v$ and $v'$ from the experimental findings [\onlinecite{dissip_neutr}]. For the real contribution $v$ to the velocity we take its low energy (up to $\sim25\mu eV$) value of $4.6\cdot 10^4 m/s$.
Concerning the imaginary part $v'$, we note that although the curve for $\mathrm{Im}(k)/\omega$ exhibits more of a quadratic behaviour we approximate it by a linear one and estimate $v' \simeq - v \mathrm{Im}[k(\omega)]/\mathrm{Re}[k(\omega)]\vert_{\omega\sim 20\mu eV}\sim -v/5$. Next, taking $|x_0| = 3\mu m$, we find the threshold energy to be of the order $\eps_{thr} \sim 40\mu eV$.
Surprisingly, using a very raw guess, we arrive at the value of the same order as in the experiment!

Returning to the general discussion, we note that two scenarios are possible -- dissipation present in one or both modes. In the former case it leads to the mere inequivalence between the two modes and thus the two contributions to the interference current do not cancel each other contrary to the case of linear dispersion \eqref{charge_asympt_simple_1}. The second possibility reveals even more interesting physics as one can demonstrate the full coherence recovery under certain conditions. Indeed, a parameter responsible for the decoherence strength\cite{artur} without dissipation is $\eps_0\eta$, in a sense that for $\eps_0\eta\gg 1$ the visibility vanishes. Therefore, ``turning on''  the dissipation the parameter becomes $\eta/\gamma$ for $\eps_0\gamma\gg 1$. Consequently, when $\eta/\gamma\ll 1$ in both modes the WP should not decohere at all, which we demonstrate in the next section. It can already be understood from the following arguments. Decoherence originates from the inelastic processes\cite{imry} inside the interferometer, whose probability increases with growing $\eta/\gamma$, as $\gamma$ can be considered as a characteristic time width of the WP. Thus, in case of a large WP broadening scattering can be described perturbatively in terms of the above parameter. Therefore, the lower value of this parameter leads to a higher position of the visibility plateau, up to a complete coherence recovery, described by the visibility $V_0$ in \eqref{visibilit_free}.
It is therefore not surprising that the case of $x_0 = 0$ (dissipation only inside the interferometer) results in the complete decoherence of the WP with increasing initial energy. On the other hand, the situation of $|x_0|\gg L$ implies a strong dissipation and a large energy loss before the interferometer. It is also the case that provides a clear physical picture and can be relatively simply analyzed analytically.

Finally, we note that the similar reasoning is applicable to the dispersion  case.
Although the energy is conserved, the energy density lowers down as a result of broadening of the WP. Hence, the strength of decoherence is described by the same ratio of $\eta$ to the associated broadening of the WP in time.
Therefore, such a threshold effect in the visibility might as well be explained by the presence of dispersion. We support this statement in Sec. \ref{sec_disp} by  finding the visibility asymptotics when a weak dispersion is present in one mode.

%An observed partial coherence recovery in case of strong dissipation can be explained by the energy loss, proportional to $1/\gamma$ (speaking about one mode), by the time the wave packet reaches the interferometer. The details on the energy loss can be found in the Appendix \ref{appendix_dissip_energy}. Thus, the energy of the WP is not enough for non-elastic processes to appear, which prevents a WP from a complete decoherence. Indeed, as it was already discussed\cite{artur}, the parameter governing the visibility behaviour in the case of just a linear spectrum is $\eps_0\eta$. Namely, when $\eps_0\eta\gg 1$ visibility vanishes. However, we see how adding dissipation modifies such condition, introducing a new energy scale $1/\gamma$. 

\section{Visibility: effect of dissipation}\label{sec_dissip}

To proceed we may directly use the expression for the interference current \eqref{interf_current}, bearing in mind that the shift $\gamma$ is not small and may be different for $F(u,v)$ and $F(v,u)$. Let us then study the current in case of $\gamma$ being much larger or much smaller compared to the flight time difference $\eta$.

\subsubsection{Strong dissipation: $\gamma\gg\eta$}\label{sec_strong_dissip}

Considering the dissipation parameter $\gamma$ to be large compared to $\eta$ is natural in our model since we demonstrated that $\gamma\propto x_0$, while $|x_0|\gg L$.
Let us first assume that dissipation is only present in the neutral mode, then
\begin{align}\label{large_dissip_int_charge_general}
&\mc{Q}= -i\frac{\tau_{\ml}\tau^{\ast}_{\mr}}{2\pi u v \eta}\int_{-\infty}^{\infty}dt \lbrace F(u,v)-F(v,u)\rbrace,\\
&F(u,v) = \frac{\sqrt{t-i\gamma}}{\sqrt{t+i\gamma}}\frac{\sqrt{t-\eta+i\gamma}}{\sqrt{t-\eta-i\gamma}},\label{large_dissip_F_neutral}
\end{align}
with finite\footnote{In fact, in the lowest order the visibility does not depend on in which mode exactly the dissipation is ``turned on''.} $\gamma$ in $F(u,v)$ and infinitesimally small in $F(v,u)$. When $\gamma\gg\eta$, we can expand $F(u,v)$ as:
\begin{equation}
F(u,v) = 1 + \frac{i\eta\gamma}{\gamma^2+t^2}.
\end{equation}
Meanwhile,  for the part, originating from the charge mode
\begin{equation}\label{F_charge_mode}
F(v,u) = \sgn(t)\sgn(t-\eta), 
\end{equation}
so that
\begin{align}
\int_{-\infty}^{\infty}dt\lbrace F(u,v)-F(v,u)\rbrace = \eta(2 + i\pi).
\end{align}
Substituting it into \eqref{large_dissip_int_charge_general} and using a definition \eqref{interf_current_charge_int} for the interference charge we get:
\begin{equation}\label{dissipation_strong_one_mode}
Q_{int} = \alpha\frac{2\vert \tau_{\ml}\tau_{\mr}\vert}{uv} \cos(\vp_{AB}+\theta),
\end{equation}
with $\alpha e^{i\theta} = 1/2-i/\pi$, i.e. $\alpha \sim 0.592$.
As we mentioned before the direct charge does not change, so we arrive at the following result for the visibility in case of strong dissipation present in the neutral mode:
\begin{equation}\label{dissipation_strong_one_mode_visib}
V = \alpha V_0.
\end{equation}

\begin{figure}[h]
\includegraphics[scale=0.35]{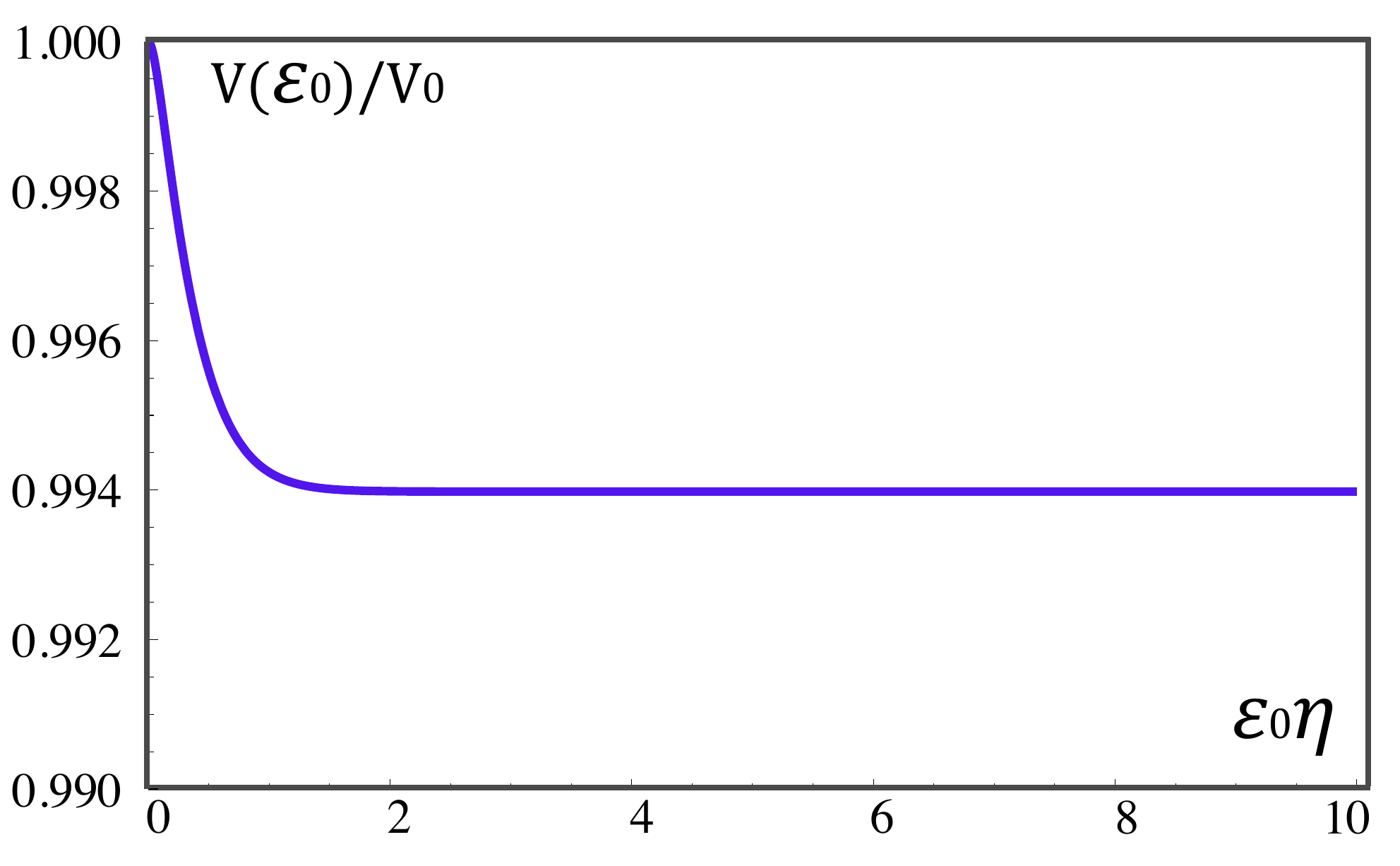} 
\caption{Visibility as a function of the initial wave-packet energy  in case of strong dissipation in both modes, calculated from \eqref{visib_strong_dissip}. Here $\gamma_1 = 2.38\eta, \gamma_2 = 5.68\eta$.}
\label{fig:visib_str_dissip}
\end{figure}

Interestingly, including strong dissipation into both modes, the coherence is completely recovered and the visibility acquires the maximum value $V_0$ in \eqref{visibilit_free}, corresponding to the free fermion case! 
Particularly, using the expression for the current \eqref{interf_current} with dissipation $\gamma_1$ for the neutral mode and $\gamma_2$ for the charge one and expanding it in terms of $\eta/\gamma_{1,2}$, we can simply perform all integrations without taking the limit $\eps_0\to\infty$. Eventually, we are able to present the visibility as a function of initial energy of the wave packet in the following form:
\begin{align}\label{visib_strong_dissip}
&V = V_0 \left(1-\frac{\eta^2}{16}A +\frac{\eta^2}{128}B\right),\\
& A =\sum_{n=1,2}\frac{1}{\gamma_n}\left(\frac{1}{2\gamma_n}-e^{-2\eps_0\gamma_n}[\eps_0+\frac{1}{2\gamma_n}]\right),\\
& B = \Big[\sum_{n=1,2}(-1)^{n-1}\frac{1}{\gamma_n}\left(1-e^{-2\eps_0\gamma_n}\right)\Big]^2.
\end{align}
Let us check simple asymptotics. When $\eps_0\to 0$ visibility is strictly $V_0$, while for $\eps_0\to\infty$ it saturates to a slightly smaller value 
\begin{equation}
V \simeq V_0\Bigg[1-\frac{\eta^2}{32}\left(\frac{1}{\gamma^2_1}+\frac{1}{\gamma^2_2}\right)+\frac{\eta^2}{128}\left(\frac{1}{\gamma_1}-\frac{1}{\gamma_2}\right)^2\Bigg].
\end{equation}
The plot of \eqref{visib_strong_dissip} is presented in Fig.\ \ref{fig:visib_str_dissip}.

\subsubsection{Weak dissipation: $\gamma\ll \eta$}

Let us study the other limit, when dissipation is small and is present in one, neutral, mode for simplicity. Again we are going to make use of the expressions for the current from the previous section \eqref{large_dissip_int_charge_general} and \eqref{F_charge_mode} and study, therefore, the $\eps_0\to\infty$ asymptotics. Note that the integral over time for the difference $F(u,v)-F(v,u)$ obviously converges, although it is not the case for each of the terms separately. Hence, we are going to rewrite it as $\lbrace F(u,v)-1\rbrace -\lbrace F(v,u)-1\rbrace$, so that each brace contains a well converging integrand over time. We denote the terms as $\mc{F}(u,v)$ and $\mc{F}(v,u)$ respectively.

\begin{figure}[h]
\includegraphics[scale=0.6]{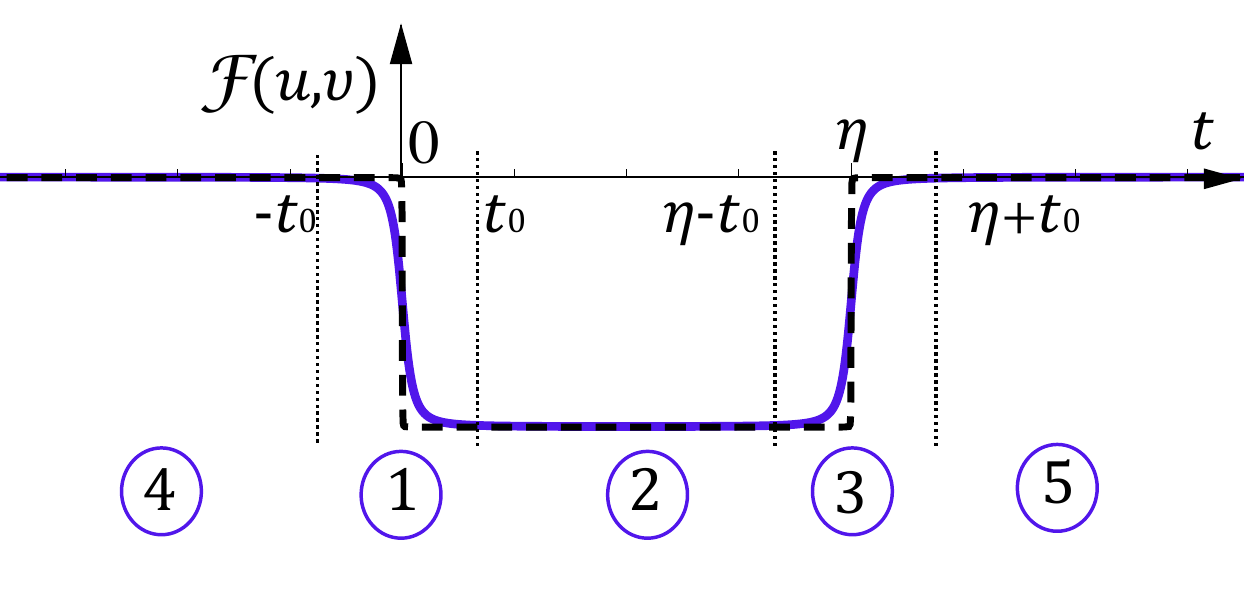} 
\caption{The plot of $\mc{F}(u,v)\equiv F(u,v)-1$ (dashed line) for $\gamma\to 0$ and its real part (solid line) in case $\gamma\ll\eta$ as a function of $t$.The region is divided into five numbered areas over which the integration is performed. Parameter $t_0$ is chosen such that $\gamma \ll t_0\ll \eta$.}
\label{fig:modes_separation} 
\end{figure}

Consider $F(v,u)$ (alternatively $F(u,v)$) in \eqref{large_dissip_F_neutral} when $\gamma\ll \eta$. Although it is already simply represented, we may rewrite it as follows for further needs:
\begin{equation}\label{F_modes_separation}
\mc{F}(u,v) = - \frac{\sqrt{t-i\gamma}}{\sqrt{t+i\gamma}} +\frac{\sqrt{t-\eta+i\gamma}}{\sqrt{t-\eta-i\gamma}}.
\end{equation}
The Fig.\ \ref{fig:modes_separation} gives an idea of how the above expression looks like as a function of $t$. Finite $\gamma$ smears sharp borders at $t=0$ and $t=\eta$ in the real part of the function.
However, the limit of small dissipation allows using of \eqref{F_modes_separation} to integrate $\mc{F}(u,v)$ in the $1$st and $3$rd regions and also  expanding \eqref{large_dissip_F_neutral} in terms of $\gamma/\eta$ on the rest of the time axis.
For that it is divided into five regions so that $\int dt \mc{F}(u,v)=\sum_{n=1}^5 J_n$, where $J_n$ is a contribution to the integral from the $n$th area. Then, using $t_0\ll \eta$  and taking into account the above explanations, the integrals over the regions in the vicinity of $t=0$ and $t=\eta$ read
\begin{align}
J_1+J_3 &= \int_{-t_0}^{t_0} dt\left(- \frac{\sqrt{t-i\gamma}}{\sqrt{t+i\gamma}} +\frac{\sqrt{t+i\gamma}}{\sqrt{t-i\gamma}}\right)\\
 &= 4i\gamma \ln \frac{2t_0}{\gamma}.
\end{align}
From the careful expansion of \eqref{large_dissip_F_neutral} in other regions we get:
\begin{align}
J_2 = 2t_0-\eta + 2i\gamma&\ln \frac{\eta}{t_0}, \quad J_4 = 4i\gamma\ln\frac{2\eta}{\gamma},\\
&J_5 = i\gamma\frac{\eta}{t_0}.
\end{align}
Hence, $\int dt \mc{F}(u,v) = 2t_0-\eta +4i\gamma\ln\frac{2\eta}{\gamma}$.
The first two terms are eliminated in the expression for the current by the contribution  $\int dt F^{\ast}(v,u) = 2t_0 -\eta$ from the charge mode. Thus, the visibility is described by the following formula:
\begin{equation}
V = V_0 \frac{2\gamma}{\pi\eta}\ln\frac{2\eta}{\gamma}.
\end{equation}

\section{Visibility: effect of dispersion}\label{sec_disp}

As we have already mentioned before dispersion effect on the visibility is similar to that of dissipation despite a qualitatively different behavior of the WP in its presence. However, the broadening of the WP in space in presence of dispersion reduces effectively the possibility of the inelastic scattering inside the interferometer, which should lead to a coherence recovery. To provide a quantitative evidence for it, we again study the limit of a small interferometer $|x_0|\gg L$ and, for simplicity, assume a quadratic dispersion to be present in a neutral mode. 

We adopt the same method used in Sec. \ref{sec_corr_func} and start by solving the equation of motion for the bosonic field:
\begin{align}
\lbrace i\omega-v_0\partial_x - (&v_1-v_0)\theta(x-x_0)\partial_x\nonumber\\
&+i\gamma\partial_x\theta(x-x_0)\partial_x\rbrace\phi(\omega,x) = 0,
\end{align}
with the dispersion of the form $i\gamma\partial_x\theta(x-x_0)\partial_x$, which corresponds to a Hermitian operator. The velocity $v_1$ takes the desired values $v$ or $u$, depending for which mode dispersion is taken into account. Unlike the dissipation case, we can directly solve this equation, using the solution for $\phi(x<x_0)\equiv\phi^{-}$, derived straightforwardly from the expression \eqref{phi_exp}, and performing the Fourier transform:
\begin{align}
\phi^{-}(k) = \int \frac{dk'}{\sqrt{k'}} \left(\hat{a}_{k'} \frac{\delta(\omega-k'v_0)}{i(k'-k)+\delta}e^{i(k'-k)x_0}\right.\nonumber\\
\left.+\hat{a}^\dagger_{k'}\frac{\delta(\omega+k'v_0)}{-i(k+k')+\delta}e^{-(k'+k)x_0}\right).
\end{align}

Considering $\gamma$ to be small, the solution for $\phi(x>x_0)\equiv\phi^{+}$ reads:
\begin{align}
\phi^{+} \simeq \int\frac{dk'}{\sqrt{k'}} \Big[\hat{a}_{k'} \exp\lbrace-ik'(v_0 t-x_0)\nonumber\\
+i\frac{v_0}{v_1}k'(x-x_0)(1+k'\gamma\frac{v_0}{v^2_1})\rbrace + h.c.\Big]
\end{align}
Having such an expansion we formally write down all the needed correlation functions and arrive at the following expression for the integral over time for the auxiliary interference charge $\mc{Q}$:
\begin{align}
\mc{Q} = -\frac{\Gamma\tau_{\ml}\tau^{\ast}_{\mr}}{2\pi^2uv\eta}\int dt\iint_{-\infty}^0 d\tau d\tau'\frac{e^{i\eps_0(\tau-\tau')+\Gamma(\tau+\tau')}}{\tau-\tau'-i\delta}\nonumber\\
\times\left(e^{\frac{1}{2}\lbrace\mc{K}_1(t,\tau)+\mc{K}_2(t,\tau')\rbrace}-F(v,u)\right),\label{dispersion_Q_1}
\end{align}
where $F(v,u)$ was defined in \eqref{interf_current_F}, while correlators $\mc{K}_1$ and $\mc{K}_2$ take dispersion into account as follows:
\begin{align}
&\mc{K}_1(t,\tau) = \int_0^{\infty}\frac{d\omega}{\omega}e^{i\omega\tau+i x_0\frac{\omega}{v}(1+\frac{\omega\gamma}{v^2})+i\omega t}(e^{i\eta\omega}-1),\\
&\mc{K}_2(t,\tau') = -\mc{K}_1^{\ast}(t,\tau').
\end{align}
Let us bring the integral \eqref{dispersion_Q_1} to a dimensionless form by making a substitution of variables
\begin{align}
\omega =\beta\tilde{\omega}, \quad t = \tilde{t}/\beta, &\quad \tau = \tilde{\tau}/\beta, \quad \tau' = \tilde{\tau}'/\beta,\\
&\beta = v \sqrt{\frac{v}{x_0\gamma}}
\end{align}
and consider a limit $\eps_0\to\infty$. For simplicity we are also studying the case $\beta\eta\ll 1$, making it analogous to the problem of a strong dissipation in one mode, discussed in Sec. \ref{sec_strong_dissip}.
We, thus, arrive at the result
\begin{align}\label{dispersion_Q_prefinal}
&\mc{Q} = \frac{i\tau_{\ml}\tau^{\ast}_{\mr}}{2\pi uv\eta}\int_{-\infty}^{\infty}\frac{d\tilde{t}}{\beta}\left(\mathrm{sgn}\left(\frac{\tilde{t}}{\beta}\right)\mathrm{sgn}\left(\frac{\tilde{t}}{\beta}-\eta\right)-e^{i\eta\beta J(\tilde{t})}\right),\\
&J(\tilde{t}) = \int_0^{\infty}d\tilde{\omega}\cos(\tilde{\omega}\tilde{t}+\tilde{\omega}^2).
\end{align}
Note that $J(\tilde{t})$ as a function of $\tilde{t}$ is well bounded and hence the exponent in \eqref{dispersion_Q_prefinal} may be expanded in terms of $\eta\beta$:
\begin{align}
&\mc{Q} = -\frac{i\tau_{\ml}\tau^{\ast}_{\mr}}{2\pi uv}\left(2+i\int_{-\infty}^{\infty}d\tilde{t}J(\tilde{t})\right).
\end{align}
The integral in the above expression can be handled with by rewriting it in a following way:
\begin{equation}
\int_{-\infty}^{\infty}d\tilde{t}J(\tilde{t}) = \int_0^{\infty}d\tilde{t}\int_{-\infty}^{\infty}d\tilde{\omega}\cos(\tilde{\omega}\tilde{t}+\tilde{\omega}^2) = \pi.
\end{equation}
Using the definitions \eqref{interf_current_charge_int} and \eqref{visibilit_def}, and taking into account that the direct charge does not change, we find the visibility
to be exactly the same as the one found in the case of strong dissipation in one mode \eqref{dissipation_strong_one_mode}-\eqref{dissipation_strong_one_mode_visib}! From such a result we may conclude that it is the energy density that plays a role in the coherence recovery effect. In case of a dissipation it decreases because of the energy loss, while for a dispersion the reason is the space broadening of the WP.

\section{Periodic coherence recovery for linear spectrum}\label{sec_period}

In Sec. \ref{visib_linear} we demonstrated how a separation of a WP into two quasiparticles in a linear spectrum case leads to visibility decay with increasing initial energy. Taking into account dissipation or dispersion effects we were able to present an explanation of partial coherence recovery within conditions of a particular experiment. On the other hand, it is of interest to imagine a new experiment which would allow to introduce ``unequal'' conditions of propagation for the two modes, so that their exact cancellation would be impossible already in a simple case of a linear spectrum.
As a nature of the effects mentioned above is not yet understood, we consider an experiment, where they can be neglected (or one could take $x_0=0$). Then, a periodic bias $\Delta\mu \sin(\Omega t)$ is applied between the upper and lower arms of the interferometer, so that $\Delta\mu\ll \Omega$. Additional phase shifts appear then in the electron operators in the upper arm of MZI and the asymptotics for the  interference current at $\eps_0\to\infty$ reads\footnote{The general shift is described by $\exp\left(i\Delta\mu/\Omega[\cos\Omega(t-L/v)-\cos\Omega t']\right)$. However, following \eqref{delta_diff}, such a phase is cancelled in front of $F(v,u)$ when $t'=t-L/v.$}
\begin{align}
&\langle I_{\ml\mr}(t)\rangle =
2 i \frac{\tau_{\ml}\tau^{\ast}_{\mr}}{2\pi uv\eta}\\
&\times\Big[
e^{i\frac{\Delta\mu}{\Omega}\lbrace\cos\Omega(t-L/v)-\cos\Omega(t-L/u)\rbrace} F(u,v)-F(v,u)
\Big]\nonumber
\end{align}
Such a result implies the periodic coherence recovery, which we will study by expanding the exponent in the above expression up to the first order in $\Delta\mu/\Omega$. However, before proceeding, we point out that any additional random phase appearing in the arguments of the cosines results in averaging out of the effect. Hence, the periodic signal applied between the arms of MZI must be synchronized with the tunneling event from the QD.

Integrating current over time, we arrive at the following expression:
\begin{align}\label{periodic_ch}
&\mc{Q} \propto \int_{-\infty}^{\infty}dt\sgn(t-\eta)\sgn t\sin\left(\Omega\Big[t-\frac{x_0}{v}-\eta\Big]+\varphi\right),\\
&\varphi = \arccos \frac{2\Gamma}{\sqrt{4\Gamma^2+\Omega^2}}.
\end{align}
There are two contributions to the current: from the WP and the ``background'', which is left after the WP has passed through the interferometer. It is periodic unlike the DC current, consisting of electrons rarely tunneling from the QD into the edge. Thus, these signals can be measured separately, which correspond to leaving integration over $t\in[0,\eta]$ in \eqref{periodic_ch}.
Finding the interference charge and taking into account that the direct charge is not influenced by the applied bias, we express the saturated visibility in the lowest order of $\Delta\mu/\Omega$ as
%\begin{equation}
%V = 4V_0\frac{\Delta\mu\cos\varphi}{\Omega}\frac{\sin^2\frac{\eta\Omega}{2}}{\eta\Omega}\sin\left(\Omega\Big[\frac{x_0}{v}+\frac{\eta}{2}\Big]-\varphi\right),
%\end{equation}
\begin{equation}
V = 4V_0\vert \Delta\mu\sin\theta\vert\frac{\sin^2\frac{\eta\Omega}{2}}{\eta\Omega^2}\cos\varphi,
\end{equation}
where $\theta = \Omega\Big[\frac{x_0}{v}+\frac{\eta}{2}\Big]-\varphi$. As expected the periodic potential leads to the imbalance between the contributions of the two quasiparticles and an oscillating visibility for $\eps_0\eta\gg 1$. Notably, in case $\eta\Omega\ll 1$ and $\eta\ll x_0$, the parameter determining the periodicity of the visibility recovery is mostly $\Omega x_0/v$, while the amplitude  becomes proportional to $\Delta\mu\eta$. Thus, the periodic signal is able to partially restore the coherence of the WP, which is a direct consequence of the fact that the charge and neutral modes behave as quasiparticle, which do not dephase on their own.

\section{Conclusions}\label{sec_concl}

In this paper we make several important statements. For a start, we provide a plausible explanation of the recent experiment [\onlinecite{main_roche}], where a robust coherence plateau in the visibility of a single electron state, initially injected into a QH edge channel, was detected. The visibility of this electron WP, sent to the MZI, was found to remain almost constant as a function of the injected energy starting from a certain threshold value, while having been expected to decay and vanish. This expectation was supported 
in the theoretical work [\onlinecite{artur}], where a bosonization approach was used to account for a strong Coulomb interaction present in the QH edge system. It made obvious the inability of such a minimalistic model to capture the effect of the coherence recovery. Analyzing the reason behind it, we find that ultimately the origin of the visibility decay lies in the destructive interference between the two quasiparticles arising from the interaction in the QH edge at the filling factor $\nu=2$. Nevertheless, each of the quasiparticles remains coherent, which is a puzzling outcome. We check that this conclusion is robust with respect to the strength of interaction, i.e. when the charges of these quasiparticles deviate from $1/2$. Thus, a trivial mechanism of the partial coherence recovery could be implemented in the model by merely creating an imbalance between the two contributions of the quasiparticles. One source of such an imbalance is an asymmetry of the interferometer. However, we show that a slight asymmetry that could be present in the experiment is not responsible for such a strong coherence recovery. Moreover, modifying the lengths of the arms of the MZI with gates it is easy to get rid of this contribution completely. 

The two other candidates, namely dissipation or dispersion, which were found to be present\cite{dissip_neutr} at least in the neutral mode, can also account for the aforementioned imbalance.
However, most importantly, they can provide the second means of the coherence recovery. Indeed, in the presence of either dissipation or dispersion the energy density of the WP is lowered by the time it reaches the interferometer. Hence, it weakens the inelastic scattering inside it, which recovers the coherence. Therefore, the stronger the energy loss results in the bigger coherence recovery. In support of this reasoning, we demonstrate that coherence can be fully recovered if a strong dissipation is present in both modes. We point out that our calculations are performed assuming the distance between the QD and the interferometer $|x_0|$ to be much larger than the size of interferometer $L$. In this limit we are able to predict the values of the coherence plateau in cases of strong/weak dissipation and also weak dispersion. Considering $|x_0|\gg L$ allows to neglect the effects of dissipation or dispersion taking place inside the interferometer, so we are not in a position to predict exact values of the visibility plateau or the threshold energy in a particular experiment. However, what we do show is that dissipation or dispersion inside the interferometer on their own ($x_0=0$) do not lead to coherence recovery. Thus, it is the outside dynamics which plays a significant role. These assumptions should be easy to check experimentally by studying the values of the visibility plateau as a function of $x_0$ that must be increasing with it. 
Moreover, the rough estimations of a threshold energy, defined only by the dissipation strength, are in a good agreement with the experiment.

Finally, we return to the question of the quasiparticle nature of the two edge excitations and propose an experiment which could shed light on their physics. Namely,  we study a set-up where a periodic bias of a small amplitude, synchronized with the tunneling event from the QD, is applied between the arms of a symmetric interferometer. Assuming a linear spectrum, one then discovers sustained oscillations of the visibility at a large injected energy, which is exactly a consequence of the intrinsic coherence of the two modes. Although, dissipation or dispersion might also manifest themselves it would be important to compare all the contributions.

{\it{\bf Acknowledgements.}}
This project was supported by the Swiss National Science Foundation.

\onecolumngrid
\appendix

\section{Correlation functions with dissipation}\label{appendix_dissip_corr}

Following the formula \eqref{dissip_field_sol_1}
and a relation between a charge density and a bosonic field, one gets the following expression for the linear response function:
\begin{align}
\mathcal{G}(k',k,k'',-\omega) = \frac{2\pi k}{v}
\left(\frac{\delta(k'-\frac{\omega}{v})\delta(k''+\frac{\omega}{v}+k)}{k+\frac{\omega}{v+iv'}}
- \frac{\delta(k'-\frac{\omega}{v})\delta(k''+\frac{\omega}{v}+k)-\delta(k+k')\delta(k'')}{k+\frac{\omega}{v}-i\delta}\right),
\end{align}
which immediately reveals
\begin{align}
S_{\rho}(k',k,k'',\omega) =-\frac{4\pi k \omega}{v\tilde{v}'} \theta(\omega)\frac{\delta(k'-\omega/v)\delta(k''+\omega/v+k)}{(k+\omega /\tilde{v})^2+\omega^2/\tilde{v}'^2}
\end{align}
with the help of the FDT statement \eqref{dissip_fdt}, where $\tilde{v}' = \frac{v^2 + v'^2}{v'}$. Hence, the spectral function acquires the following form in the coordinate space, valid for $v' < 0$:
\begin{align}\label{app_S}
S_{\rho}(y,x,x_0,\omega) = -\frac{\omega(\tilde{v}^{-1}+i\tilde{v}'^{-1})}{2\pi v}e^{i\frac{\omega}{v}(y-x_0)+i\omega(\tilde{v}^{-1}+i\tilde{v}'^{-1})(x_0-x)}.
\end{align}
Integrating now this expression over the coordinates and taking care of the boundary conditions, we arrive at the correlator for the fields:
\begin{align}
S_{\phi}(y,x,x_0,\omega) = 4\pi^2\int_{-\infty}^{y} dy'\int_{x}^{+\infty} dx' S_{\rho}(y',x',x_0,\omega) =\frac{2\pi}{\omega}e^{-i\frac{\omega}{v}(x_0-y)-i\omega (x-x_0)(\tilde{v}^{-1}+i\tilde{v}'^{-1})}.
\end{align}
The seeming divergence at zero $\omega$ is non-physical and disappears in the correlators for the electron operators. Performing a Fourier transform we find the expression \eqref{dissip_correlator}. A simple check in the last expression shows that in the limit $\tilde{v}' = 0$, we restore the correlation function for free bosonic fields.

\section{Energy of a mode with dissipation}\label{app_energy_dissip}

Here we find the energy $E_v$ of the neutral mode with dissipation, following the discussion of the energy distribution in Sec. \ref{sec_energy_distr_dissip} and using the expression \eqref{energy_distr_neutral} in the limit $\eps_0\to\infty$:
\begin{align}
E_v = \int d\eps f_v(\eps)\eps  \propto \int dt \int dz \frac{\delta'(z)}{z-i\delta}\left(\frac{\sqrt{t+i\gamma}\sqrt{t+z-i\gamma}}{\sqrt{t-i\gamma}\sqrt{t+z+i\gamma}}-1\right) = i\int dt \frac{2t\gamma-i\gamma^2}{2(t^2+\gamma^2)^2}=\frac{\pi}{4\gamma},
\end{align}
where we used integration by parts. Restoring all the coefficients $E_v = \frac{1}{16\pi}\sqrt{\frac{v}{u}}\gamma^{-1}$. Note that taking the limit $\gamma\to 0$ we recover the energy ultraviolet cut-off, as it should be. Energy for the charge mode is found exactly in the same manner. 

\section{Decoherence in case of $x_0=0$.}\label{appendix_x0}

To include dissipation inside the interferometer into consideration we must modify the expression for the current \eqref{interf_current}. Again we are going to study its integral over time $\mc{Q}$, from which the interference charge is found. Then from the general definition  \eqref{current_def} in case $x_0=0$:
\begin{equation}
\mc{Q} \equiv\int_{-\infty}^{\infty} dt \langle I_{\ml\mr}(t)\rangle = \int_{-\infty}^{\infty} dt \int_{-\infty}^t dt' \left(M(t,t')+M(t',t)\right),
\end{equation}
where 
\begin{align}
M(t,t') \propto \frac{\sqrt{t-L/u+i\gamma_2}}{\sqrt{t-L/u-i\gamma_2}}\frac{\sqrt{t-L/v+i\gamma_1}}{\sqrt{t-L/v-i\gamma_1}}\frac{t'-i\delta}{t'+i\delta}\left(\frac{\gamma_1}{(t'-t+L/v)^2+\gamma^2_1}-\frac{\gamma_2}{(t'-t+L/u)^2+\gamma^2_2}\right),
\end{align}
with $\delta$ being infinitesimally small and $\gamma_{1,2}$ being finite to account for the dissipation. Note that when $\gamma_{1,2}$ are also infinitesimally small the lorentzians in the above expressions become Dirac $\delta$-functions. Therefore, as the integration for $t'$ comes from the region $t' \leqslant t$, the contribution from $M(t',t)$ vanishes. In this manner the expression \eqref{interf_current} can be found. However, with dissipation being present inside the interferometer, $M(t',t)$ needs to be taken into account. We first rewrite the integrals over $t'$, making them running up to $t'=0$. 
Next, we shift $t$ by $-t'$ and change the sign of $t'$ in $M(t',t)$, which allows to rewrite $\mc{Q}$ as an integral over the whole axis of $t'$:
\begin{align}
\mc{Q} \propto \int_{-\infty}^{\infty} dt\frac{\sqrt{t-L/u+i\gamma_2}}{\sqrt{t-L/u-i\gamma_2}}\frac{\sqrt{t-L/v+i\gamma_1}}{\sqrt{t-L/v-i\gamma_1}}  \int_{-\infty}^{\infty}dt' \frac{t+t'-i\delta}{t+t'+i\delta}\left(\frac{\gamma_1}{(t'+L/v)^2+\gamma^2_1}-\frac{\gamma_2}{(t'+L/u)^2+\gamma^2_2}\right).
\end{align}
Integration over $t'$ can be easily performed by moving the integration contour in the positive half plane, which results in zero for the whole expression.

\section{Interference current in the limit $\eps_0\to \infty$ for arbitrary fractional charges of the quasiparticles}\label{appendix_fract}

The integral over time of the interference current at a large initial energy $\eps_0$ acquires the form:
\begin{align}\label{app_curr_fract_1}
\mathcal{Q }= \int_{-\infty}^{\infty}dt\langle I_{\mr\ml}(t)\rangle \propto\int_{-\infty}^{\infty}dt\int_{-\infty}^{\infty} d\tilde{t} \chi_v(t,\tilde{t})\chi_u(t,\tilde{t})\mathcal{F}(\tilde{t}),
\end{align}
where 
\begin{align}
\chi_v(t,\tilde{t}) = \frac{\left(t-\tilde{t}+\frac{x_0}{v}-i\gamma\right)^{\delta_1}}{\left(t-\tilde{t}+\frac{x_0}{v}+i\gamma\right)^{\delta_1}}\frac{\left(t+\frac{x_0-L}{v}+i\gamma\right)^{\delta_1}}{\left(t+\frac{x_0-L}{v}-i\gamma\right)^{\delta_1}}
\end{align}
and $\chi_u(t,\tilde{t})$ is defined in the same manner by replacing $v\to u$ and $\delta_1\to\delta_2$.
The last multiplier in \eqref{app_curr_fract_1} is defined as follows:
\begin{align}\label{app_F}
\mathcal{F}(\tilde{t}) = \frac{1}{\left(i(\tilde{t}-L/v)+\gamma\right)^{2\delta_1}\left( i(\tilde{t}-L/u)+\gamma\right)^{2\delta_2}}
-\frac{1}{\left(-i(\tilde{t}-L/v)+\gamma\right)^{2\delta_1}\left(-i(\tilde{t}-L/u)+\gamma\right)^{2\delta_2}}.
\end{align}
If $\delta_1 = \delta_2 = 1/2$, then 
\begin{equation}\label{delta_diff}
\mc{F} \propto \delta(\tilde{t}-L/v) - \delta(\tilde{t}-L/u),
\end{equation}
so that we easily arrive at the previous formulas for the interference charge  \eqref{charge_asympt_simple_1}, which eventually results in  zero for the asymptotics. When the charges are arbitrary, $\mathcal{F}$ can not be  simplified. However, we are still able to derive a few conclusions. Firstly,  $\int_{-\infty} ^{\infty}d\tilde{t}\mathcal{F}=0$. It can be justified by the fact that 
for $\tilde{t}$ outside the region $[L/u,L/v]$ the integrand quickly decays, due to the  $1/t^2$ decrease of both terms in $\mathcal{F}$. Hence, as both branch points in each term are situated in the same half-plane, the contour can be modified by being dragged into the infinity of the other half-plane. Such a procedure shows that the integral indeed vanishes. 

Thus, we can subtract the unity from $\chi_v\chi_u$ in \eqref{app_curr_fract_1}. The advantage is that $\chi_v\chi_u - 1$ becomes non-zero only in certain regions of $t$. Obviously, these regions are defined by various $\tilde{t}$, but the only relevant $\tilde{t}$s belong to $[L/u,L/v]$, as we realized above from analyzing $\mathcal{F}$. In this case the value of $\chi_v\chi_u$ can be simply found:
\begin{equation}
\chi_v\chi_u-1=e^{2i\pi\delta_1}-1,  
\end{equation}
for $t\in [\frac{L-x_0}{u},\tilde{t}-\frac{x_0}{u}]\cup [\tilde{t}-\frac{x_0}{v},\frac{L-x_0}{v}]$ and zero everywhere else. 
Therefore, integrating $\chi_v\chi_u-1$ over the indicated region in $t$ shows that there is no a dependence on $\tilde{t}$, so we are left with $\mathcal{Q}\propto \int d\tilde{t}\mathcal{F} = 0$. 

\section{Interference charge in case of a small asymmetry in the MZI for a linear dispersion}\label{app_asymm}

Adding an asymmetry to the interferometer (the upper arm of the length $L_1$ and the lower one of $L_2\gtrsim L_1$) one arrives at the expression for the interference charge of a form resembling the one \eqref{app_curr_fract_1} in the previous Appendix:
\begin{align}
\mc{Q} \sim \int_{-\infty}^{\infty} dt \int_{-\infty}^{\infty} d\tilde{t} \chi_v(t,\tilde{t})\chi_u(t,\tilde{t})\mc{F}(\tilde{t}),
\end{align}
with $\chi_{v,u}(t,\tilde{t})$ describing the propagation of the neutral or charge modes in the upper arm of the interferometer
\begin{equation}
\chi_v(t,\tilde{t}) = \sgn\left(t+\frac{x_0-L_1}{v}\right)\sgn\left(t-\tilde{t}+\frac{x_0}{v}\right),
\end{equation}
and  $\mc{F}(\tdt)$ being responsible for the interference effect:
\begin{equation}\label{F_asymm}
\mc{F}(\tdt) = \frac{1}{\{(\tdt-L_1/u-i\delta)(\tdt-L_2/u-i\delta)(\tdt-L_1/v-i\delta)(\tdt-L_2/v-i\delta)\}^{1/2}} - c.c.,
\end{equation}
where $\delta$ is an infinitesimally small shift.
In case of a symmetric interferometer $L_1 = L_2$ the above expression $\mc{F}(\tdt)$ acquires a simple form \eqref{delta_diff} in terms of the delta-functions. 

To find where the main contribution to $\mc{F}(\tdt)$ comes from, we multiply  and divide the second term in it by the denominator of the first one, which immediately results in the simplification
\begin{equation}
\mc{F}(\tdt) =\frac{1-\sgn(\tdt-L_1/u)\sgn(\tdt-L_1/v)\sgn(\tdt-L_2/v)\sgn(\tdt-L_1/v)}{\{(\tdt-L_1/u-i\delta)(\tdt-L_2/u-i\delta)(\tdt-L_1/v-i\delta)(\tdt-L_2/v-i\delta)\}^{1/2}}.
\end{equation}
It now becomes clear from the form of the numerator that the only relevant $\tdt$ lies in the following region: $\tdt \in [L_1/u;L_2/u] \cup [L_1/v;L_2/v]$. Here we have assumed that $L_1/v \gg L_2/u$, which seems natural due to the fact that $u\gg v$ and that we are mainly interested in a case of a small difference between $L_2$ and $L_1$. Moreover, the integral of $\mc{F}(\tdt)$ over $\tdt$ is simply zero similarly to the statement made in the previous Appendix. Hence, we may rewrite $\mc{Q}$ as
\begin{equation}
\mc{Q} \sim \iint dt d\tdt \big[\chi_u(t,\tdt)\chi_v(t,\tdt)-1\big] \mc{F}(\tdt).
\end{equation}
The idea is then to study the expression in the square brackets as a function of $t$, using the particular values for $\tdt$. Let us denote firstly $t_1 = \frac{L_1-x_0}{u}$, $t_2 = \frac{L_1-x_0}{v}$, $t_3 = \tdt - \frac{x_0}{u}$, $t_4 = \tdt -\frac{x_0}{v}$. Next, it is easy to show that
\begin{align}
\text{for}\ \tdt\in [L_1/u;L_2/u]:\ \chi_u(t,\tdt)\chi_v(t,\tdt) = 
\begin{cases}
-2, \text{if}\ t\in [t_1,t_3] \cup [t_4,t_2],\\
0,\ \text{elsewhere}
\end{cases}\\
\text{for}\ \tdt\in [L_1/v;L_2/v]:\ \chi_u(t,\tdt)\chi_v(t,\tdt) = 
\begin{cases}
-2, \text{if}\ t\in [t_1,t_3] \cup [t_2,t_4],\\
0,\ \text{elsewhere},
\end{cases}
\end{align}
where the last expression holds for $\frac{L_2-L_1}{v}<x_0\left(1/u-1/v\right)$, which essentially means that the difference $L_2-L_1$ must be smaller than $x_0$ and is quite applicable to the conditions of the experiment. With that in mind, we can perform integration over $t$ and arrive at the following expression 
\begin{align}
\mc{Q} \sim \eta_1 \mc{I}_1 + \mc{I}_2;\quad \mc{I}_1 = \int _{L_1/u}^{L_2/u}d\tdt \mc{F}(\tdt),\ \text{and}\ \mc{I}_2 = \int_{L_1/v}^{L_2/v}d\tdt\left(2\tdt - L_1[1/u+1/v]\right) \mc{F}(\tdt).
\end{align}
with the notation $\eta_1 = L_1(1/v-1/u)$ similar to the one used throughout the main part of the paper. 

Both $\mc{I}_1$ and $\mc{I}_2$ are table integrals\cite{gradshteyn2007} if rewritten in a suitable form. Indeed, we express the first integral $\mc{I}_1$ as follows
\begin{equation}
\mc{I}_1 = -2i\int_{L_1/u}^{L_2/u} \frac{d\tdt}{\{(\tdt-L_1/u)(L_2/u-\tdt)(L_1/v-\tdt)(L_2/v-\tdt)\}^{1/2}} = -4i \frac{F(\frac{\pi}{2}, r)}{\sqrt{L_1L_2}(1/v-1/u)},
\end{equation}
where $F(\pi/2,r)\equiv \int_0^{\pi/2} \frac{d\alpha}{\sqrt{1-r^2\sin^2\alpha}}$ is an elliptic integral of the first kind and $r = \frac{L_1-L_2}{\sqrt{uvL_1L_2}(1/u-1/v)}$. Note that, when $L_1 = L_2$, then $r = 0$ and $F(\pi/2,0)=\pi/2$ , while $\eta_1\mc{I}_1(L_1=L_2) = -2\pi i$.
We deal with the second integral in the same manner
\begin{align}
\mc{I}_2 &= 2i\int_{L_1/v}^{L_2/v} d\tdt\frac{2\tdt -L_1(1/u+1/v)}{\{(\tdt-L_1/u)(\tdt-L_2/u)(\tdt-L_1/v)(L_2/v-\tdt)\}^{1/2}}\nonumber\\
&= \frac{4i}{\sqrt{L_1L_2}(1/v-1/u)}\Bigg[2\left(\frac{L_2}{v}-\frac{L_1}{u}\right)\Pi\left(\frac{\pi}{2}, \frac{L_1-L_2}{v\eta_1},r\right)+\frac{L_1}{u} F\left(\frac{\pi}{2},r\right) - L_1\left(\frac{1}{u}+\frac{1}{v}\right)F\left(\frac{\pi}{2},r\right)\Bigg],
\end{align}
where $\Pi\left(\frac{\pi}{2}, \frac{L_1-L_2}{v\eta_1},r\right) \equiv \int_0^{\pi/2}\frac{d\theta}{\left(1-\frac{L_1-L_2}{v\eta_1}\sin^2\theta\right)\sqrt{1-r^2\sin^2\theta}}$ is an elliptic integral of the third kind. Again a simple check for $L_1=L_2$ gives $\Pi\left(\frac{\pi}{2}, \frac{L_1-L_2}{v\eta_1}, 0\right) = \pi/2$, which results in $\mc{I}_2(L_1 = L_2) = 2\pi i$. Therefore, the total charge $\mc{Q} = 0$ in this limit as it should be. Combining the above findings for $\mc{I}_1$ and $\mc{I}_2$ we finally get:
\begin{equation}\label{int_charge_asymm}
\mc{Q} \sim \frac{i}{\sqrt{L_1L_2}(1/v-1/u)}\Bigg[\left(\frac{L_2}{v}-\frac{L_1}{u}\right)\Pi\left(\frac{\pi}{2}, \frac{L_1-L_2}{v\eta_1},r\right)-\eta_1F\left(\frac{\pi}{2},r\right)\Bigg].
\end{equation}
The experiment [\onlinecite{main_roche}] was stated to be conducted with a symmetric interferometer. Thus, accounting for only a slight asymmetry and studying the expression \eqref{int_charge_asymm} as a function of $L_2$ in the vicinity of $L_1$
one arrives at the interference charge of the order $Q \sim \frac{L_2-L1}{v\eta_1}$. Therefore, an asymmetry indeed leads to a partial coherence recovery, but apparently it can not be the main reason behind its large value in the experiment.
%already visually judging by the parameters of the interferometer.

\twocolumngrid

\bibliographystyle{apsrev4-1}
\bibliography{coherence_recovery}

\end{document}